# Modeling Growth Curve of Fractal Dimension of Urban Form of Beijing

Yanguang Chen; Linshan Huang

(Department of Geography, College of Urban and Environmental Sciences, Peking University, Beijing 100871, P.R.China. E-mail: chenyg@pku.edu.cn)

**Abstract**: The growth curves of fractal dimension of urban form take on squashing effect and can be described by sigmoid functions. The fractal dimension growth of urban form in western countries can be modeled by Boltzmann's equation and logistic function. However, these models cannot be well applied to the fractal dimension growth curve of Beijing city, the national capital of China. In this paper, the experimental method is employed to find parametric models for the growth curves of fractal dimension of Chinese urban form. By statistical analysis, numerical analysis, and comparative analysis, we find that the quadratic Boltzmann equation and quadratic logistic function can be used to characterize how the fractal dimension of the urban land-use pattern of Beijing increases in the course of time. The models are also suitable for many cities in the north of China. In order to convert the empirical models into theoretical models, we attempt to construct a model of spatial replacement dynamics of urban evolution, from which the logistic model of urban fractal dimension growth can be derived. The models can be utilized to predict the rate and upper limitation of Chinese urban growth. In particular, the models can be employed to reveal the similarities and differences between the fractal growth of Chinese cities and that of the cities in western countries.

# 1. Introduction

Fractal geometry is one of effective means of describing complex systems and exploring complexity. The analysis of urban patterns and processes suggests new ways of understanding spatial complexity (Allen, 1997; Batty, 2005; Batty, 2013; Wilson, 2000). One of the basic properties of complex systems is scale invariance, which indicates that the scale-free systems cannot be described with the conventional mathematical methods based on characteristic scales. Fractal geometry is one of powerful tools for exploring complex spatial systems such as cities (Batty and Longley, 1994; Chen, 2008; Frankhauser, 1994). Urban evolution involves urban form (pattern) and growth (process). Urban form can be characterized by fractal dimension, and urban growth can be reflected by fractal dimension change. On the other, urban form is one of the important aspects of urbanization (Knox and Marston, 2009). The change of the level of urbanization over time takes on a sigmoid curve, which is termed *urbanization curve* (Cadwallader, 1996; Pacione, 2005). Urbanization curves are often formulated as a logistic function (Chen, 2009; Karmeshu, 1988; United Nation, 1980). This suggests that the fractal dimension growth curve of urban form can be characterized by some kind of sigmoid function. A discovery is that fractal dimension growth of urban form in developed countries and regions can be modeled with Boltzmann's equation, which can be reduced to a logistic function (Chen, 2012; Chen, 2014a; Chen, 2018). However, if we use the ordinary Boltzmann's equation or logistic function to describe the growth curves of fractal dimension of urban form in China, the effect is not satisfying in many cases. The models cannot be well fitted to the observational time series of fractal dimension of Chinese urban form.

Spatial measurements show that the fractal dimension growth of urban form of Beijing city does take on sigmoid curves. However, there are subtle differences between the fractal dimension growth curves of Beijing's urban form and those of the cities in developed countries such as London and Baltimore. By a lot of times of empirical analyses, we find that the growth curves of fractal dimension of Beijing's urban form can be modeled by a generalized logistic function rather than the conventional logistic function. In fact, the proper model of fractal dimension growth curves of Beijing's urban form is a quadratic logistic function, which can be regarded as a special case of quadratic Boltzmann's equation. This finding is meaningful because it suggests that the spatial dynamics of urban evolution in China maybe differs from that in western developed countries. In fact, the urbanization curves of both China and India can be described with the quadratic logistic function instead of the usual logistic function (Chen, 2016). The process of urbanization may influence a city's growth and form in a certain way (Knox and Marston, 2009). Thus the new models of fractal dimension growth curves may be generalized to the cities in less developed countries. Based on experimental method, this paper is devoted to making parametric models of fractal dimension growth curves of urban form in China. In Section 2, a quadratic Boltzmann's equation is proposed to describe the fractal dimension growth curve of urban form of Beijing. For normalized fractal dimension, quadratic Boltzmann's equation can be simplified as a quadratic logistic function. A spatial dynamics model is presented to explain the quadratic logistic process. In Section 3, the quadratic sigmoid functions are applied to the typical multifractal parameters of Beijing city. In Section 4, the related questions are discussed. Finally, the discussion is concluded by summarizing the main points of this study.

## 2. Models

### 2.1 The idea of model construction

Cities represent a type of complex spatial systems, which bear a set of service functions to the surrounding areas. There used to be three key concepts about city, that is, *city proper* (CP), *urbanized area* (UA), and *metropolitan area* (MA) (Davis, 1978). Today, the second concept, urbanized area, tends to be replaced by the concept of urban agglomerations in the literature on fractal cities. To study a city, we should find a set of measures or parameters to describe it. Anyway, scientific studies should proceed first by describing how a system works and later by understanding why (Gordon, 2005). The main mean of quantitative description is mathematics and measurements (Henry, 2002). Measurement is the basic link between mathematics and empirical studies (Taylor, 1983). The conventional measures include length, area, volume, and density. To describe a city, we should know its urban area, which is the precondition of measuring urban population size. Unfortunately, urban area cannot be objectively measured because of scale dependence of urban land use. In other words, urban form has no characteristic scales and cannot be effectively described by conventional mathematical methods. In this case, fractal dimension of urban form can be employed to replace urban area to reflect the degree of space filling in a city.

If we have a time series of fractal dimension of a city's form, how can we predict urban growth and analyze the spatial dynamics of the city? An effective approach is to make a mathematical model of fractal dimension growth curves. Consequently, the measurement description of urban form will develop to mathematical description of urban growth. The key issue is to choose a mathematical function that models properly the growth curve of fractal dimension. In scientific research, two methods can be used to establish mathematical models: *analytical method* and *experimental method* (Zhao and Zhan, 1991). The analytical method is based on certain theoretical principles and logic deduction, while the experimental method is based on observational data and statistical analysis. The two different methods result in two different mathematical models: *mechanistic models* and *parametric models* (Su, 1988). The former is also termed structural models, belonging to theoretical models, while the latter can be termed functional models, belonging to empirical models. An empirical model will become a theoretical model if it can be mathematically demonstrated or derived out from one or more postulates. In this work, we will utilize experimental method to build parameter models for fractal dimension growth curves (Table 1). Then, we will try to derive the empirical models from a spatial replacement principle so that they will become theoretical models.

Two concepts are important for understanding fractals and fractal dimension of cities. One is topological dimension $d_T$, and the other is the Euclidean dimension of the embedding space $d_E$. In theory, the Lebesgue measures of real fractals are zero. This suggests that a city fractal can be treated as a fractal point set, and thus the topological dimension of city fractals is $d_T = 0$. On the other hand, a city fractal can be defined in a 2-dimensional embedding space and $d_E = 2$, and it can also be defined in a 3-dimensional embedding space and $d_E = 3$ (Thomas *et al*, 2012). Majority of fractal cities are defined in a 2-dimension embedding space based on digital maps or remote sensing images (e.g. Batty and Longley, 1994; Benguigui *et al*, 2000; Feng and Chen, 2010; Frankhauser, 1994; Shen, 2002). However, some scholars studied fractal cities through 3-dimension embedding space (e.g., Qin *et al*, 2015). In fact, a fractal defined in the 3-dimensional embedding space can be explored through the 2-dimensional embedding space (Vicsek, 1989). A rational city fractal should be defined in the 2-dimensional space due to the following reasons. First, fractal dimension is used to replace 2-dimensional urban area rather than 3-dimensional urban volume. Second, the effective skill of scientific quantitative analysis is to reduce dimension instead of to increase dimension. Third, the allometric scaling relation between city population and urban area suggests that urban form should be defined in a 2-dimensional space. Generally speaking, fractal dimension of cities will come between the topological dimension $d_T = 0$ and the embedding dimension $d_E = 2$.

The fractal dimension range between the topological dimension and the embedding dimension gives rise to a concept, namely, *squashing effect*. If a variable reflects an endless process of growth, and the variable has strict upper limit (e.g., $d_E=2$ or 3) and lower limit (e.g., $d_T=0$ or 1), the track of the variable will be twisted into a S shape. The S-shaped curve can be described by a sigmoid function, which is also termed squashing function (Mitchell, 1997). In fact, there are a number of sigmoid functions that can be employed to characterize squashing curves. The family of sigmoid functions include conventional logistic function, generalized logistic function, arc-tangent function, hyperbolic tangent function, Gompertz function, Boltzmann equation, and generalized Boltzmann equation. To select a proper function to model the fractal dimension growth curves of urban forms, we should make the best of logic judgments and statistical analysis. Comprehensive comparison and analysis show that the Boltzmann equation and logistic function can be used to describe the fractal dimension growth of the cities in western countries (Chen, 2012; Chen, 2018). However, these functions cannot be directly applied to the growth curves of fractal dimension of many Chinese cities such as Beijing. To characterize the tracks of fractal dimension growth, we must improve Boltzmann's equation and logistic function. A hypothesis is that quadratic Boltzmann equation and quadratic logistic function are suitable for Chinese cities. The key points are tabulated to make clearer the idea and process of model building (Table 1).

**Table 1 The main points of parametric modeling of the growth curves of fractal dimension of urban form**

| Item | Content | Explanation |
| --- | --- | --- |
| Purpose | Modeling | Find proper functions to model fractal dimension growth curves of urban form |
| Logical basis | Squashing effect | A endless growing variable confined by strict lower and upper limits forms a S-shaped curve |
| Available functions | Sigmoid function family | Logistic function, arc-tangent function, hyperbolic tangent function, Gompertz function, Boltzmann equation, etc. |
| Experiment | Verification | Use numerical, statistical, and comparative analyses to test and calibrate models |

| Selection | Criterion | (1) Statistics: goodness of fit ($R^2$); (2) Logic: $D_{max} \leq d_E = 2$. |
|---|---|---|
| Function | Models' uses | (1) Predict urban growth; (2) Explain urban evolution; (3) Sharpen urban questions |

## 2.2 Quadratic models of fractal dimension growth curves

A growth curve of fractal dimension is a smooth trend line of the time series of fractal dimension values. This type of curves bears an analogy with the urbanization curves and maybe it can be termed "fractal dimension curves" (Chen, 2018). Squashing effect is significant in the fractal dimension growth curves of urban form. First, a fractal dimension growth curve reflects the continuous process of urban growth; second, the fractal dimension growth curve can be naturally extended in the direction of time; third, the fractal dimension growth curve is subject to the definite upper limit and lower limit. The lower limit, or the minimum dimension, $D_{min}$, is determined by the topological dimension of urban form, $d_T$, and the upper limit, or the maximum dimension, $D_{max}$, is determined by the Euclidean dimension of the embedding space of city fractals, $d_E$. Among various squashing functions, the most possible function for describing the sigmoid curves of fractal dimension growth is Boltzmann's equation. The equation was originally proposed in physics, and can be used to model urban phenomena (Benguigui et al, 2001). In fact, a mathematical model in one scientific field can be usually applied to another scientific field, and the same model can be used to describe multiple phenomena in many different fields. However, in a certain field, a model parameter has its own unique meanings. As a mathematical model, Boltzmann's equation possesses four necessary parameters, the upper limit ($D_{max}$), lower limit ($D_{min}$), initial value ($D_{(0)}$), and original rate of growth ($k$) of fractal dimension. Boltzmann's equation can depict the S-shaped curves, and indeed can be used to model the fractal dimension increase of the cities in western countries (Chen, 2012; Chen, 2018). Unfortunately, when we utilize the equation to describe the fractal dimension growth curves of Beijing city, the result is not good. On the one hand, the model cannot be well fitted to the observational data of Beijing's fractal dimension series; on the other hand, we cannot find an effective convergent value for the maximum fractal dimension, $D_{max}$, which must be less than the embedding dimension, $d_E$. A hypothesis is that the conventional Boltzmann equation should be substituted by the quadratic Boltzmann equation. The revised Boltzmann model is as below:

$$D(t) = D_{\min} + \frac{D_{\max} - D_{\min}}{1 + [\frac{D_{\max} - D_{(0)}}{D_{(0)} - D_{\min}}]e^{-(kt)^2}} = D_{\min} + \frac{D_{\max} - D_{\min}}{1 + \exp(-\frac{t^2 - t_0^2}{p^2})}, \quad (1)$$

in which $D(t)$ denotes the fractal dimension of urban form in time of $t$, $D_{(0)}$ is the fractal dimension in the initial time, $D_{\max} \leq d_E$ refers to the upper limit of fractal dimension, $D_{\min} \geq d_T$ refers to the lower limit of fractal dimension, $k$ is the inherent growth rate, $p$ is the temporal scaling parameter, and $t_0$, a temporal translational scale parameter indicating a critical time scale. The scale and scaling parameters can be respectively expressed as $p=1/k$ and $t_0 = p[\ln((D_{\max}-D_{(0)})/(D_{(0)}-D_{\min}))]^{1/2}$. For the normalized fractal dimension, equation (1) can be reduced to a quadratic logistic function

$$D^*(t) = \frac{D(t) - D_{\min}}{D_{\max} - D_{\min}} = \frac{1}{1 + (1/D_{(0)}^* - 1)e^{-(kt)^2}}, \quad (2)$$

where $D^*(t)$ denotes the fractal dimension normalized by the range, and $D_{(0)}^* = (D_{(0)} - D_{\min})/(D_{\max} - D_{\min})$ indicates the normalized result of the original fractal dimension value, $D_{(0)}$. The fractal dimension range is the difference between the maximum dimension and the minimum dimension, i.e., $D_{\max} - D_{\min}$. In theory, we have $D_{\max} = d_E = 2$ and $D_{\min} = d_T = 0$ (Thomas et al, 2007). In practice, however, the numerical ranges of parameters are $1 < D_{\max} \leq d_E = 2$ and $d_T = 0 \leq D_{\min} \leq 1$. In fact, the box-counting method can be employed to estimate the fractal dimension values of urban form. There are two approaches to making use of the box-counting method. One is the method of fixed maximum box based on constant study area, and the other is that of unfixed maximum box based on variable study area. For the former case, fractal dimension values come between 0 and 2; while for the latter case, fractal dimension values vary from 1 to 2 (Chen, 2012).

The model of the 1-dimensional dynamics can be derived from the logistic model of the fractional dimension growth curves of urban form. The derivative of equation (2) is a quadratic logistic equation

$$\frac{dD^*(t)}{dt} = 2k^2 t D^*(t)[1 - D^*(t)], \quad (3)$$

which reflects the growth rate of fractal dimension. Without loss of generality, the time interval can be set as $\Delta t=1$. Thus, equation (3) can be discretized to yield a 1-dimensional map such as

$$D_{t+1}^* = (1 + 2k^2 t) D_t^* - 2k^2 t D_t^{*2}. \quad (4)$$

Using equation (4), we can generate a time series of fractal dimension for urban form, which is consistent with the values predicted by equation (2). For the comparability of fractal dimension values, the same maximum box can be applied to different study areas in different years. In this case, the minimum dimension value of growing city fractals can be treated as $D_{\min}=d_T=0$ in theory. Thus, equation (1) can be reduced to

$$D(t) = \frac{D_{\max}}{1+(D_{\max}/D_{(0)}-1)e^{-(kt)^2}}, \tag{5}$$

which is a quadratic logistic function (Chen, 2016). The normalized fractal dimension can be simplified as $D_t^*=D(t)/D_{\max}$. Accordingly, equation (4) can be rewritten as

$$D_{t+1} = (1+2k^2t)D_t - \frac{2k^2t}{D_{\max}}D_t^2, \tag{6}$$

which is equivalent to equation (4) but can be used to simulate fractal dimension growth. In other words, equation (6) is a 1-dimensional map based on quadratic logistic function and non-normalized fractal dimension. This map differs from the well-known logistic map, which can generate complex dynamics such as periodic oscillations and chaos (May, 1976). The corresponding differential equation of equation (6) is as follows

$$\frac{dD(t)}{dt} = 2k^2tD(t)[1-\frac{D(t)}{D_{\max}}], \tag{7}$$

which is equivalent to equation (3) and reflects the velocity of fractal dimension growth. It is easy to demonstrate that equation (5) is just the special solution to equation (7). In other words, equation (5) is a quadratic logistic equation of fractal dimension growth curves of urban form.

## 2.3 Tentative derivation of quadratic logistic model

In the studies on fractal cities, fractal dimension as a characteristic parameter of scale-free distributions is actually a substitute of urban area, which is a scale-dependent measure. In order to derive the parametric models of fractal dimension growth curves of urban form, we should define a set of measurements of cities which can be used to link fractal dimension to urban area (Appendix 1). In theory, a region can divided into urban places and rural places. However, there is no clear boundaries between urban land and rural land. If we define a study area around a city, urban land and rural land are interlaced with each other. Consequently, urban places and rural places form complicated patterns and cascade structure, which can be described with multifractal scaling (Chen, 2016). For the convenience of description and measurement, we should substitute the concepts of space-filled place and space-saved place for the notions of urban place and rural place (Chen, 2012). The logistic function of fractal dimension growth is a spatial replacement equation that represents the progressive replacement of natural areas by urban and rural built-up areas in the course of time (Chen2014a). On the base of Boltzmann's equation of fractal dimension growth, a spatial replacement index has been defined by means of the normalized fractal dimension (Chen, 2012). The replacement measurement can be further researched in the new framework of quadratic Boltzmann's equation. From equation (2), it follows that

$$\ln[\frac{D^*(t)}{1-D^*(t)}] = \ln(\frac{D_0^*}{1-D_0^*}) + (kt)^2 \,, \tag{8}$$

in which $0<D^*<1$. A spatial *filled-unfilled ratio* (FUR) of urban growth can be defined as

$$O(t) = \frac{U(t)}{V(t)} = \frac{D^*(t)}{1-D^*(t)} = \frac{D_0^*}{1-D_0^*} \exp[(kt)^2]\,, \tag{9}$$

where $U$ denotes the filled extent of urban space (e.g., space-filling area) with various artificial buildings, which can be measured by the pixel number of built-up land on digital maps, and $V$, refers to the unfilled extent of space (e.g., space-saving area), in which there is no constructions or artificial structures. Thus we have

$$D^*(t) = \frac{O(t)}{O(t)+1} = \frac{U(t)}{U(t)+V(t)} = \frac{U(t)}{S(t)}\,, \tag{10}$$

where $S$ represents the total space of urbanized region, that is, $S=U+V$. This indicates that $V$ is the complementary set of $U$, and *vice versa*. The higher the $O$ value is, the higher the extent of space-filling will be. The normalized fractal dimension can be regarded as *level of space filling* (SFL) of cities, indicating the degree of spatial replacement. It can be employed to estimate the remaining space of urban growth. The space-filling extent, $U$, is usually a known number because it can be measured using digital maps. However, the space-saving area, $V$, is often unknown (Chen, 2012). In this case, the unfilled space of an urban region can be estimated by fractal dimension values. From equation (9) it follows

$$V(t) = U(t)(\frac{1}{D^*(t)} - 1)\,. \tag{11}$$

This implies that the space-saving extent can be figured out by means of normalized fractal dimension and space-filling area. The sum of space-filling extent and space-saving extent gives the total spatial area of a fractal city, that is

$$S(t) = U(t) + V(t) = \frac{U(t)}{D^*(t)},\tag{12}$$

which is equivalent to equation (10). Equations (11) and (12) suggest a set of spatial measurements of urban growth and form (Appendix 1).

The space-filling extent and space-saving extent are both dynamic quantities. However, the total space area of a city can be treated in two different ways. One is a dynamic way, and the other, a static way. Equation (12) shows a dynamic concept for urban total space, which changes over time. In fact, a city cannot grow without any limits. There exists an ultimate spatial sphere for urban expansion. The greatest extreme of urban space can be termed *spatial capacity* of a city. Urban land use can be modeled with logistic function or quadratic logistic function (Chen, 2014b). Thus the spatial capacity of urban growth can be estimated by general logistic models. In other words, the capacity parameter of a logistic model or a quadratic logistic model can serve for the static total quantity of urban space. The fractal dimension series of a city's form can be estimated based on either fixed study area or variable study area. Based on a variable study area, the total space of urban growth can be treated as a dynamic concept; while based on the fixed study area, the urban total space should be treated as a static quantity measured by a spatial capacity parameter.

To derive the models of fractal dimension growth curves, we can make a model of 2-dimensional spatial dynamics. The process of urban growth is a nonlinear dynamic process of urban space filling and replacement. Suppose that an urban region is divided into two types: one is filled space, and the other, unfilled space. The former can measured by space-filling extent, $U(t)$, while the latter can be measured by space-saving extent, $V(t)$. Thus, the interaction between filled space (used space) and unfilled space (remaining space) can be described with two differential equations

$$\frac{dU(t)}{dt} = t[\alpha U(t) + \beta \frac{U(t)V(t)}{U(t)+V(t)}],\tag{13}$$

$$\frac{dV(t)}{dt} = t[\lambda V(t) - \beta \frac{U(t)V(t)}{U(t)+V(t)}],\tag{14}$$

where $\alpha$, $\beta$, and $\lambda$ are three parameters. This implies that the growth rate of filled space, $dU(t)/dt$, is proportional to the size of filled space, $U(t)$, and the coupling interaction between filled and unfilled space, but not directly related to unfilled space size; the growth rate of unfilled space, d$V(t)$/d$t$, is proportional to the size of unfilled space, $V(t)$, and the coupling action between filled and unfilled space, but not directly related to filled space size. The growth of unfilled space proceeds from natural land, rural area, old city transformation, counter urbanization, and so on. Differing from the spatial replacement of the cities in western developed countries (Chen, 2012), both d$U(t)$/d$t$ and d$V(t)$/d$t$ here are proportional to the product of time $t$ and $U(t)$ or $V(t)$ and the product of $t$ and $U(t)V(t)/(U(t)+V(t))$. From equations (13) and (14), we can derive equation (3). In fact, taking derivative of equation (10), we have

$$\frac{dD^*(t)}{dt} = \frac{dU(t)/dt}{U(t)+V(t)} - \frac{U(t)}{[U(t)+V(t)]^2}\left[\frac{dU(t)}{dt} + \frac{dV(t)}{dt}\right]. \tag{15}$$

Substituting equations (13) and (14) into equation (15) yields

$$\frac{dD^*(t)}{tdt} = \frac{\beta U(t)V(t)}{[U(t)+V(t)]^2} + \frac{U(t)}{U(t)+V(t)}\left(\alpha - \frac{\alpha U(t)}{U(t)+V(t)} - \frac{\lambda V(t)}{U(t)+V(t)}\right). \tag{16}$$

According to equation (10), $D^*=U/(U+V)$, $1-D^*=V/(U+V)$. So equation (16) can be expressed as

$$\frac{dD^*(t)}{dt} = (\alpha + \beta - \lambda)tD^*(t)[1 - D^*(t)]. \tag{17}$$

Comparing equations (17) with equation (3) shows

$$2k^2 = \alpha + \beta - \lambda. \tag{18}$$

This indicates that equations (13) and (14) are mathematically equivalent to equation (3). The dynamical analysis based on the 1-dimensional logistic equation can be associated with the dynamical analysis based on the 2-dimensional space replacement model. Discretizing equations (13) and (14) yields a 2-dimensional map as below

$$U_{t+1} = U_t + t(\alpha U_t + \beta \frac{U_t V_t}{U_t + V_t}), \tag{19}$$

$$V_{t+1} = V_t + t(\lambda V_t - \beta \frac{U_t V_t}{U_t + V_t}), \tag{20}$$

where $U_t$ and $V_t$ are the discrete expressions of $U(t)$ and $V(t)$, respectively. This suggests that a 1-dimensional map, equation (4), can be converted into a 2-dimensional map, equations (19) and (20). Using equations (19) and (20), we can generate a time series of approximate fractal dimension of urban growth, which is consistent with the values created by equation (4). By developing the expression of 2-dimensional spatial dynamic equations, we can further derive the quadratic Boltzmann equation of fractal dimension growth curves in this similar way.

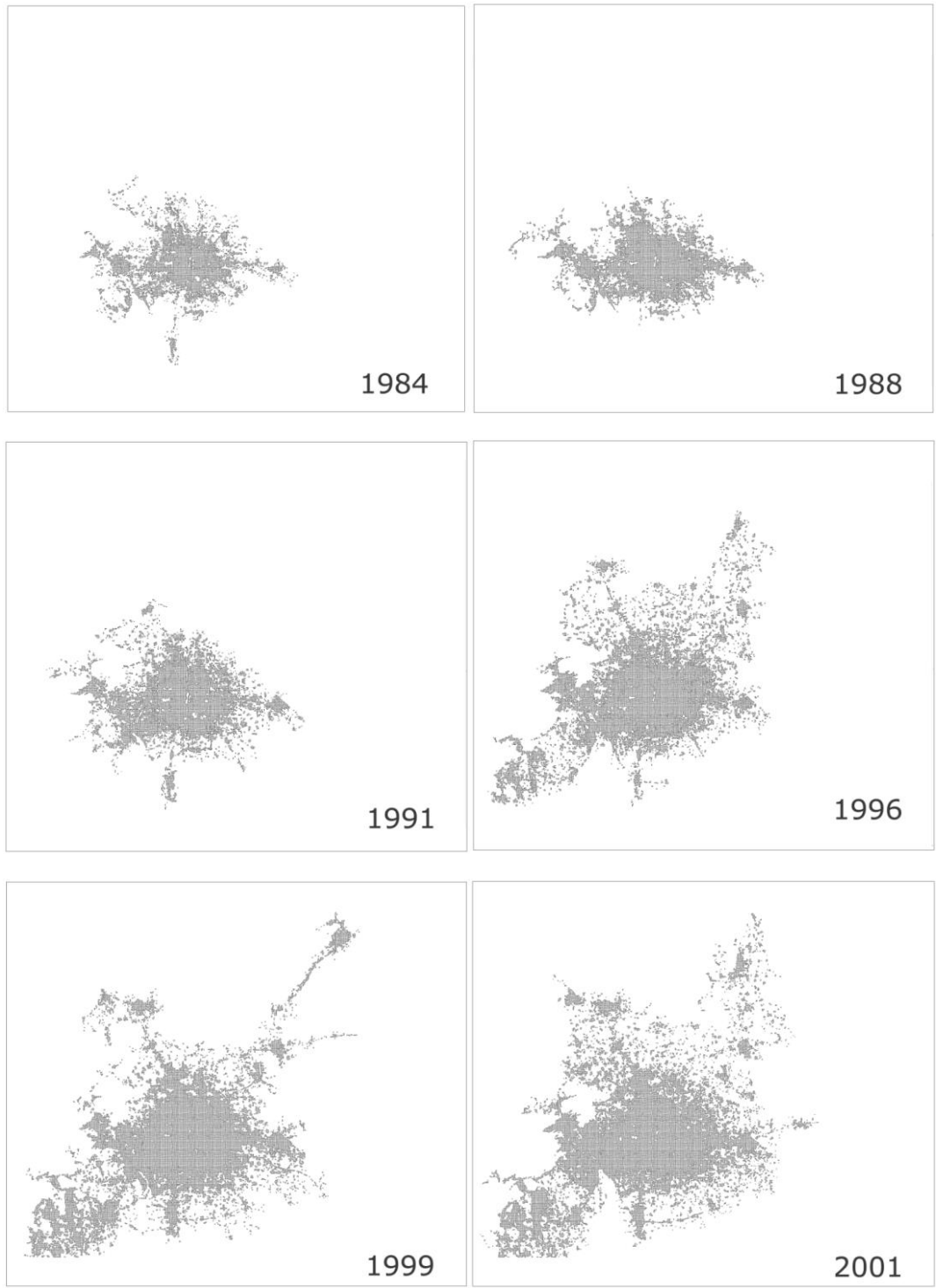

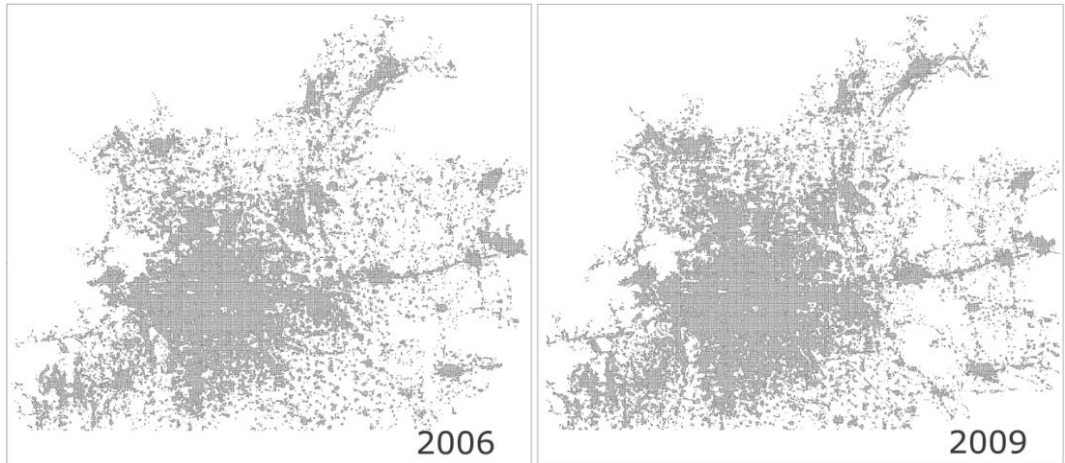

**Figure 1 Eight representative images reflecting the growth and morphology of Beijing city (1984-2009)**

**Note**: The urban a boundaries are delineated by using CCA (Rozenfeld *et al*, 2008; Rozenfeld *et al*, 2011).

## 3. Empirical analysis

### 3.1 Study area and methods

The effect of the models can be testified by the observational data of fractal dimension of urban form. The purpose of modeling is insight—polarizing thinking and sharpening questions--rather than number (Hamming, 1962; Karlin, 1983). However, the academic progress of our understanding the world relies heavily on the interplay of quantifiable data with models (Louf and Barthelemy, 2014). Therefore, before discussing the insight from a research, we must examine the goodness of fit of the models to the observed data by statistical analysis, and further, we should check the logic consistency between models and data by numerical analysis. The city of Beijing can be taken as an example to make an empirical analysis and numerical experiment. Beijing is a typical megacity with urban population more than 15 million inside the metropolitan area in 2010 (the sixth census year). The multifractal parameters of urban form can be calculated through remote sensing images for 13 years between 1984 and 2009. In fact, a number of remote sensing images of Beijing from National Aeronautics and Space Administration (NASA) are available for spatial analysis. The ground resolution of these images is 30 meters. The detailed information about these images has been clarified in previous work (Chen and Wang, 2013). The urban agglomeration of Beijing takes on complicated and irregular spatial patterns (Figure 1). The multifractal features and properties of Beijing's urban form have been illustrated and demonstrated (Chen and Wang, 2013; Huang and Chen, 2018). If we obtain a sample path of fractal dimension of urban growth in different years, the quadratic sigmoid functions can be fitted to the fractal dimension data to verify the models. Then, the modeling result can be applied to analyzing Beijing's urban growth and form.

The ArcGIS technology is employed to extract spatial data from remote sensing images, and Matlab-based computer programming is used to calculate multifractal parameters. The analytical process is as follows. **The first step is to define an urban boundary for each year**. The urban boundary is termed *urban envelope* (Batty and Longley, 1994; Longley *et al*, 1991). Several effective approaches to identifying urban boundaries have been developed recent years (Jiang and Jia, 2011; Rozenfeld *et al*, 2011; Tannier *et al*, 2011). In this paper, the "City Clustering Algorithm" (CCA) developed by Rozenfeld *et al* (2008, 2011) is employed to delineate the boundaries of Beijing's urban agglomeration in different years. By the urban envelope, we can compute urbanized area and estimate city size. **The second step is to define a study area for fractal dimension measurement.** As indicated above, there are two ways of determining study region for fractal dimension measurement (Chen, 2012). One is to fix the study area for all years of images (Batty and Longley, 1994; Chen and Wang, 2013; Shen, 2002), and the other is to select variable study area according to city size and urbanized area (Benguigui *et al*, 2000; Feng and Chen, 2010). In this study, for comparability, we define a fixed study area for the 13 years according to the urban envelope in the last year (2009). **The third step is to extract spatial data using box-counting method.** The functional box-counting method can be used to extract spatial data and calculate fractal dimension. This method is proposed by Lovejoy *et al* (1987) and consolidated by Chen (1995). It is an efficient approach to estimating fractal parameters of urban form and urban systems (Chen, 2014c; Chen and Wang, 2013; Feng and Chen, 2010; Huang and Chen, 2018). The fixed largest box is applied to the urban images in different years. Based on the maximum box, a hierarchy of functional boxes can be generated by the rule of spatial subdivision (Chen and Wang, 2013). **The fourth step is to calculate multifractal parameters using the least squares method.** By means of least squares calculation, we can determine the $f(\alpha)$ singularity spectrum using the approach developed by Chhabra and Jensen (1989) and Chhabra *et al* (1989). In order to guarantee proper multifractal spectrums, the intercepts of logarithmic linear regression models based on fractal relations are fixed to zero (Huang and Chen, 2018). Then, by Legendre's transform, we can reckon the generalized correlation dimension $D_q$. In the global multifractal dimension spectrum, capacity dimension ($D_0$), information dimension ($D_1$), and correlation dimension ($D_2$) represent three typical parameters of multi-scaling fractal form. The three parameters will be employed to verify the models for fractal dimension growth curves (Table 2).

It is necessary to clarify two problems involved with data extraction and parameter estimation. First, the quality of remoted sensing images influences fractal dimension measurement. Among the 13 years of images, only three Landsat TM images and two Landsat ETM+ images are most appropriate for fractal study because these images were chiefly taken in autumn or winter, in which there is less obstruction from clouds and plants (Chen and Wang, 2013). In order to extend the sample path of the time series of fractal dimension, we make use of all the 13 years of images we could get. The quality difference in different years causes fluctuation of fractal dimension values and urban area values. However, the random fluctuation results in variability instead of bias because positive and negative errors always counteract each other. Second, CCA is essentially a spatial search method based on spatial cluster, and different searching radius (distance threshold) results in different urban boundaries. Larger searching radius leads to larger study area. In this study, the searching radius is set as 1000 meters according to the idea from characteristic scales, which will be clarified in a companion paper. Fortunately, the two problems have impact on parameter estimation, but have no significant influence on the model expressions and statistic inference.

Table 2 Three typical multifractal parameters of Beijing's urban form and the corresponding urban area (1984-2009)

| Year | Fractal dimension | | | Urban area |
| --- | --- | --- | --- | --- |
| | Capacity dimension $D_0$ | Information dimension $D_1$ | Correlation dimension $D_2$ | $A$ |
| **1984** | 1.4994 | 1.3961 | 1.3553 | 393.3022 |
| **1988** | 1.5091 | 1.4234 | 1.3935 | 538.0889 |
| **1989** | 1.5374 | 1.4467 | 1.4105 | 530.2158 |
| **1991** | 1.5645 | 1.4726 | 1.4392 | 644.8084 |
| **1992** | 1.5849 | 1.5078 | 1.4797 | 729.0497 |
| **1994** | 1.5944 | 1.5213 | 1.4920 | 867.1342 |
| **1995** | 1.6522 | 1.5707 | 1.5373 | 1032.6239 |
| **1996** | 1.6789 | 1.5934 | 1.5560 | 1073.5342 |
| **1998** | 1.6702 | 1.5871 | 1.5531 | 1087.2600 |
| **1999** | 1.7174 | 1.6379 | 1.6056 | 1417.9022 |
| **2001** | 1.7423 | 1.6633 | 1.6308 | 1576.4091 |
| **2006** | 1.8559 | 1.7811 | 1.7462 | 2349.7967 |
| **2009** | 1.8667 | 1.7961 | 1.7647 | 2660.2626 |

**Note**: The urban area corresponds to the fractal dimension, which is calculated on the basis of urban boundaries determined by CCA.

## 3.2 Calculations

The improved model based on Boltzmann's equation is a delicate expression of the function of fractal dimension growth curves. If the box-counting method based on fixed study area is used to estimate fractal dimension of urban form, the lower limit of fractal dimension can be treated as $D_{min}=0$. In this case, Boltzmann's equation will change to a logistic function. In order to examine the growth models of the fractal dimension series, we propose a generalized Boltzmann function. This is a five parameter model indicative of sigmoid curves, which can be expressed as

$$D(t) = D_{min} + \frac{D_{max} - D_{min}}{1 + ae^{-(kt)^b}}, \qquad (21)$$

where $a=(D_{max}-D_{min})/(D_{(0)}-D_{min})-1$ is a proportionality parameter, and $b$ is a latent scaling factor. The other symbols are the same as those in equation (1). Generally speaking, the $b$ value comes between 1 and 2. If $b=1$, then equation (21) returns to the ordinary Boltzmann equation; If $b=2$, then equation (21) changes to a quadratic Boltzmann equation. If $D_{min}=0$ as given, equation (21) will become a generalized logistic function, and thus $a=D_{max}/D_{(0)}-1$. In scientific research, two methods can be used to reveal the structure and meaning of the parameters in a mathematical model: one is to find the initial value by letting $t=0$, the other is to take the derivative of the model. The former is applicable to proportionality coefficient, e.g., $a$, and the latter is suitable for characteristic parameters, e.g., $k$. Moreover, the lower limit value of fractal dimension, $D_{min}$, is determined by experience. Based on constant study area and fixed largest box, the lower limit can be treated as $D_{min}=0$; based on variable study area and unfixed largest box, the lower limit can be treated as $D_{min}=1$. If we take $D_{min}=1$, we have a quadratic Boltzmann model, while if we take $D_{min}=0$, we have a quadratic logistic model. The fractal dimension values in Table 2 is based on a fixed study area. So, the value $D_{min}=0$ is more reasonable in practice. Despite this, we fit the quadratic Boltzmann equation to the fractal dimension data of Beijing city for reference. By least squares calculation, we estimate the parameter values of both quadratic Boltzmann equation and quadratic logistic function, and the results are displayed in Table 3.

Table 3 The model parameters of fractal dimension growth curves and the goodness of fit on Beijing's urban growth (1984-2009)

| Model | Parameter /Statistic | Models of fractal dimension growth curves | | |
|---|---|---|---|---|
| | | Capacity dimension $D_0$ | Information dimension $D_1$ | Correlation dimension $D_2$ |
| Quadratic Boltzmann models | $D_{max}$ | 1.8952 | 1.8192 | 1.7866 |
| | $D_{min}$ | 1 | 1 | 1 |
| | $k$ | 0.0726 | 0.0758 | 0.0765 |
| | $a$ | 0.7851 | 0.9933 | 1.079 |
| | $b$ | 2 | 2 | 2 |
| | $R^2$ | 0.9823 | 0.9858 | 0.9875 |
| Quadratic logistic model | $D_{max}$ | 1.9171 | 1.8431 | 1.8097 |
| | $D_{min}$ | 0 | 0 | 0 |
| | $k$ | 0.0626 | 0.0639 | 0.0642 |
| | $a$ | 0.2778 | 0.3088 | 0.3173 |
| | $b$ | 2 | 2 | 2 |
| | $R^2$ | 0.9811 | 0.9846 | 0.9870 |
| Logistic model | $D_{max}$ | 2 | 2 | 2 |
| | $D_{min}$ | 0 | 0 | 0 |
| | $k$ | 0.0679 | 0.0580 | 0.0545 |
| | $a$ | 0.4302 | 0.5126 | 0.5458 |
| | $b$ | 1 | 1 | 1 |
| | $R^2$ | 0.9362 | 0.9587 | 0.9670 |
| Exponential model | $D_{max}$ | ∞ | ∞ | ∞ |
| | $D_{min}$ | 0 | 0 | 0 |
| | $k$ | 0.0099 | 0.0111 | 0.0114 |
| | $a=D_{(0)}$ | 1.4691 | 1.3765 | 1.3420 |
| | $b$ | 1 | 1 | 1 |
| | $R^2$ | 0.9723 | 0.9818 | 0.9853 |

**Note**: All these model parameter values and statistics can be verified by using the data shown in Table 2. In Boltzmann models, $a=(D_{max}-D_{(0)})/(D_{(0)}-D_{min})$, while in logistic models, $a=(D_{max}-D_{(0)})/D_{(0)}$. The latter is the special case of the former. The expressions and values of $a$ can be obtained by assuming time $t=0$ in equations (1) and (5).

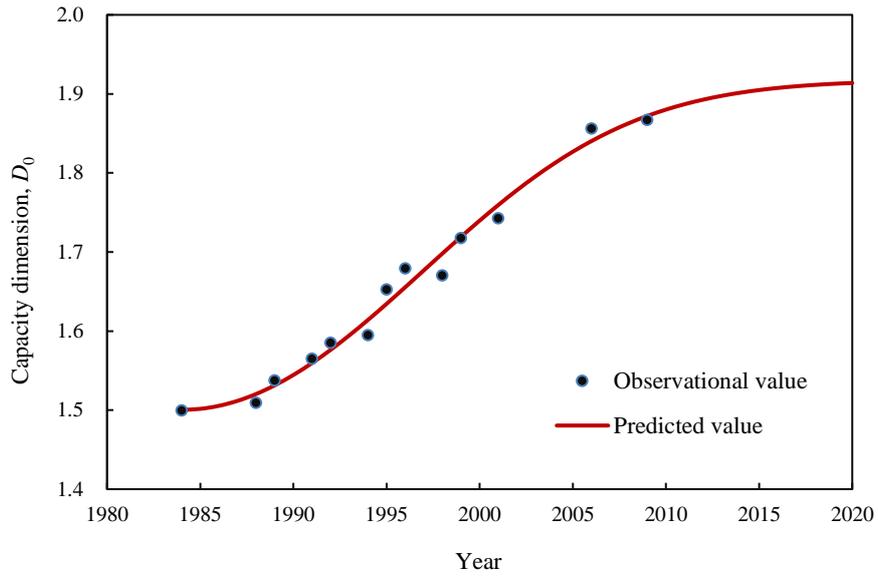

a. Capacity dimension ($1.4 < D_0 < 2.0$)

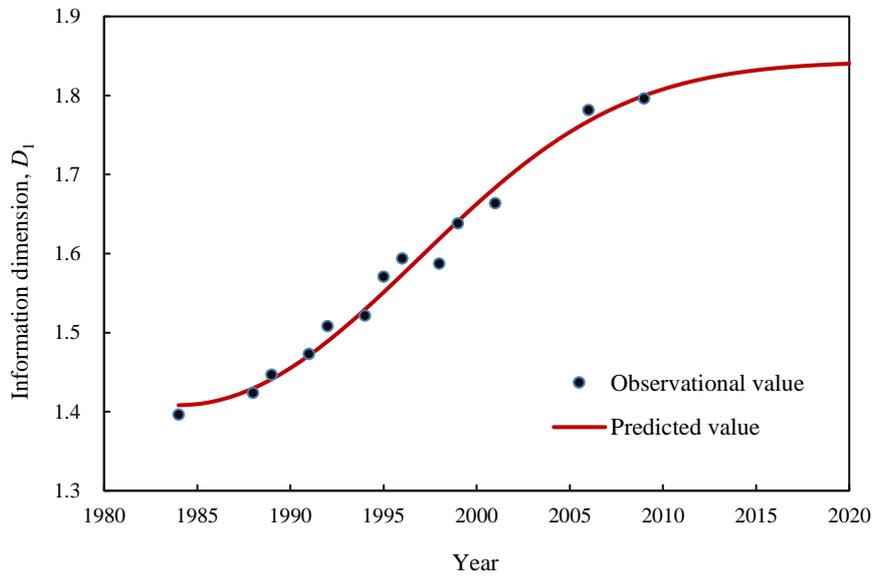

b. Information dimension ($1.3 < D_1 < 1.9$)

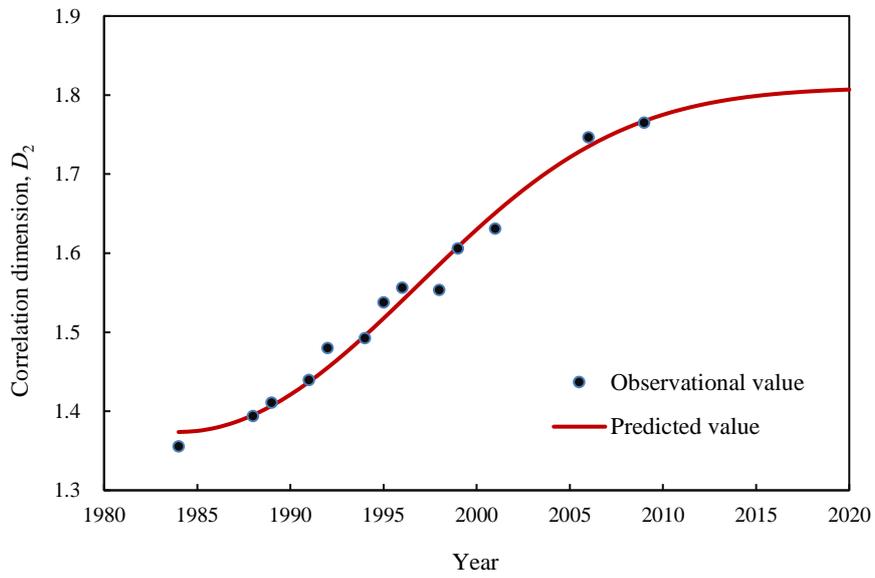

c. Correlation dimension ($1.3 < D_2 < 1.9$)

**Figure 2 Three quadratic logistic curves of fractal dimension growth of Beijing's urban form (1984-2020)**

**Note**: The trend lines are generated by using the quadratic logistic models of fractal dimension growth. The matching effect between the observed values of the fractal dimension of Beijing's urban form and the corresponding predicted values based on quadratic Boltzmann's equation is similar to the these plots.

The mathematical expressions of the quadratic growth models can be given using the above estimated parameter values. For capacity dimension ($q=0$), information dimension ($q=1$), and correlation dimension ($q=2$), the quadratic logistic models are as follows

$$\hat{D}_0(t) = \frac{1.9171}{1+0.2778e^{-(0.0626t)^2}}, \hat{D}_1(t) = \frac{1.8431}{1+0.3088e^{-(0.0639t)^2}}, \hat{D}_2(t) = \frac{1.8097}{1+0.3173e^{-(0.0642t)^2}}.$$

The goodness of fit are about $R^2=0.9811$, 0.9846, and 0.9870, respectively. The hat "^" indicates predicted value. For reference, the quadratic Boltzmann models are given as below:

$$\hat{D}_0(t) = 1 + \frac{0.8952}{1+0.7851e^{-(0.0726t)^2}}, \hat{D}_1(t) = 1 + \frac{0.8192}{1+0.9933e^{-(0.0758t)^2}},$$

$$\hat{D}_2(t) = 1 + \frac{0.7866}{1+1.0790e^{-(0.0765t)^2}}.$$

The goodness of fit are around $R^2=0.9823$, 0.9858, and 0.9875, respectively. By means of these models, we can compute the predicted value of the three types of fractal dimension. Based on the quadratic logistic models, the observed values and the calculated values of fractal dimensions can be effectively matched (Figure 2). Based on the Boltzmann model, the fitting effect is similar. The fluctuation of fractal dimension values is owing to the random disturbance proceeding from the quality of remoted sensing images, as indicated above. As far as the prediction analysis is concerned, the two models are all acceptable. However, if we consider a longer time scale, we will prefer to the logistic model. The reason is that, based on the fixed study area, the initial fractal dimension is close to $D_{min}=0$ instead of $D_{min}=1$ (Shen, 2002).

In order to compare the modeling effect of different functions, we might as well fit the general logistic model to the observational data of Beijing's fractal dimension values. If we take $D_{min}=0$ and $b=1$ in equation (21), we will have an ordinary logistic model of fractal dimension growth curves. However, the conventional logistic function cannot be well fitted to the observational data of fractal dimension of Beijing's urban form. **(1) Statistical criterion: goodness of fit.** If we take the capacity parameters $D_{max}=2$ for the $D_0$, $D_1$, $D_2$, the goodness of fit is 0.9362, 0.9587, and 0.9670, respectively. **(2) Logic criterion: capacity parameter value.** If we want to improve the goodness of fit, we must increase the capacity parameter $D_{max}$ values once and again. Unfortunately, the increase in numerical value is hundreds and thousands, and there is no end to it. This means that the capacity parameters of the conventional logistic model cannot converge to normal values. This is ridiculous, because the capacity parameter values should not exceed the Euclidean dimension of the embedding space of city fractals, that is, $D_{max} \leq d_E = 2$. This suggests that, based on the best goodness of fit, the logistic model should be replaced by an exponential growth model. It is easy to prove this inference. If $b=1$ and $D_{max} \to \infty$ as given, then equation (21) will become

$$D(t) = \frac{1}{1/D_{max} + (1/D_{(0)} - 1/D_{max})e^{-kt}} = D_{(0)}e^{kt}, \tag{22}$$

in which $1/D_{max} \to 0$. This is an exponential function. More extremely, if fractal dimension grows very fast in the short term, the exponential function will be substituted by a hyperbolic function. In light of Taylor series expansion, approximately, we have

$$\frac{1}{D(t)} = \frac{1}{D_{(0)}} e^{-kt} \approx \frac{1}{D_{(0)}} - \frac{k}{D_{(0)}} t. \tag{23}$$

which is a type of hyperbolic function.

The exponential function and hyperbolic function represent two extreme forms of the logistic function. The parameters values and the goodness of fit of the logistic function and exponential function are estimated by the least squares method and listed in Table 3 for comparison. Readers can try to test the fitting effect of hyperbolic functions and other sigmoid functions. Due to the limited space of the article, we will not give the calculation results one by one. It needs to be made clear that the choice of empirical models cannot rely solely on the goodness of fit. The exponential model and hyperbolic functions can be well fitted to the fractal dimension series of Beijing, but the two models are not acceptable. Using this kind of models to predict fractal dimension growth, fractal dimension values will quickly break the upper limit determined by the embedding dimension. All in all, the selection of models should take into account both the statistical analysis effect and logical rationality.

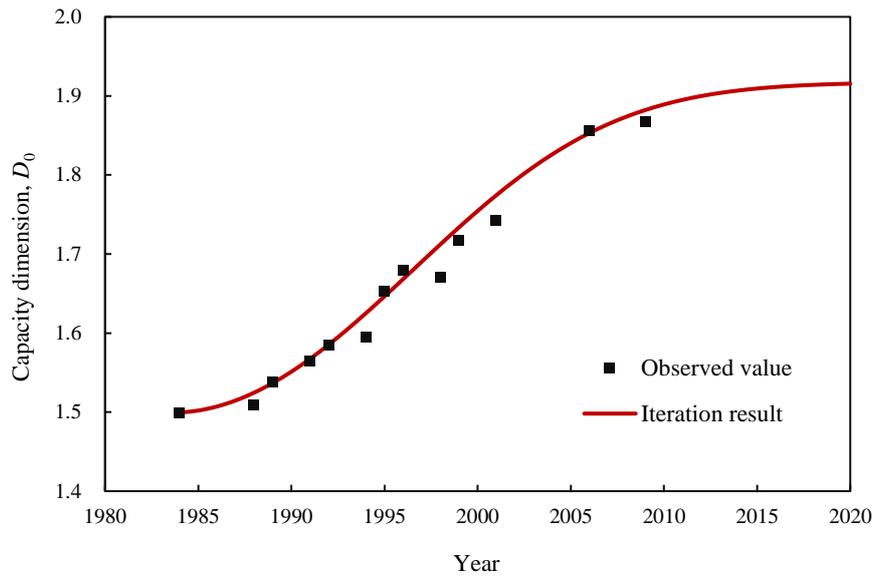

a. Capacity dimension ($1.4 < D_0 < 2.0$)

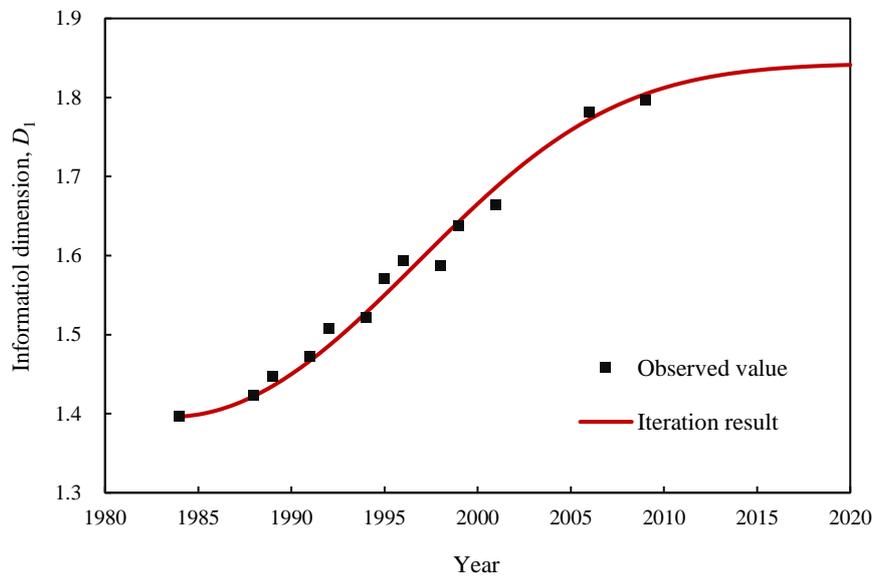

b. Information dimension ($1.3 < D_1 < 1.9$)

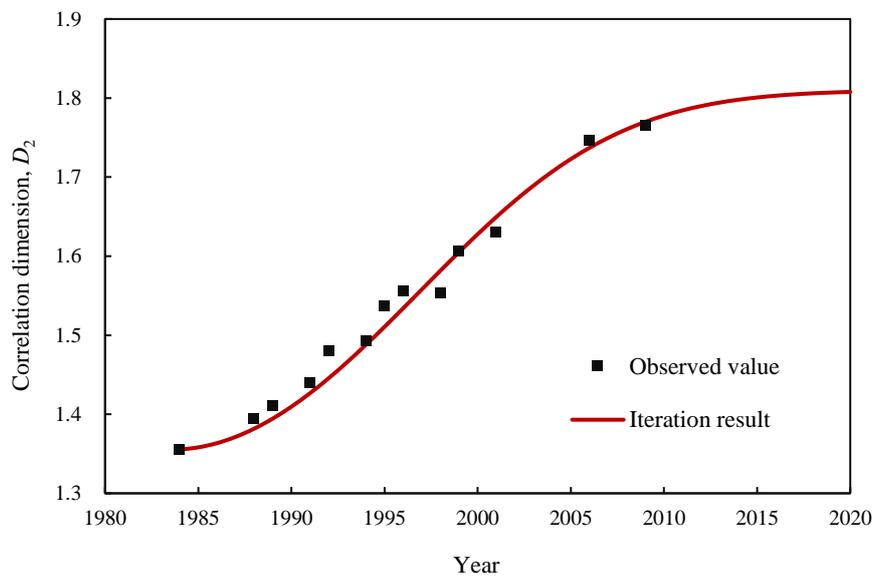

c. Correlation dimension ($1.3<D_2<1.9$)

**Figure 3 Three types of fractal dimension values and the corresponding iteration results using 1-dimensional map (1984-2020)**

**Note**: The trend lines are generated by using the 1-dimensional quadratic logistic map of fractal dimension growth.

## 3.3 Numerical iteration analysis

The quadratic logistic map can be utilized to make an analysis of numerical iteration for Beijing's urban growth. This iteration is in fact the simplest numerical simulation of fractal dimension growth. Numerical analysis can testify the modeling effect from the computational angle of view. Equations (4) and (6) are 1-dimensional maps based on the quadratic logistic function. Equation (6) can be used to simulate the fractal dimension growth, and equation (4) can be used to simulate the normalized fractal dimension growth. The two numerical calculation processes are equivalent to one another, and we just need to investigate one of them. Using a 1-dimensional quadratic logistic map, we can generate the time series of multifractal parameters of Beijing's urban form. Suppose that the initial time is 1984, and time number is $t=n-1984$, where $n=1984, 1985, …, 2020$ denotes year. For capacity dimension, the initial value $D_{(0)}=D_{(1984)}=1.4994$ (see Table 2), the maximum value $D_{max}=1.9171$ (see Table 3); For information dimension, the initial value $D_{(0)}=D_{(1984)}=1.3961$ (see Table 2), the maximum value $D_{max}=1.8431$ (see Table 3); For correlation dimension, the initial value $D_{(0)}=D_{(1984)}=1.3553$ (see Table 2), the maximum value $D_{max}=1.8097$ (see Table 3). By means of mathematical software such as Matlab or even spreadsheet such as MS Excel, we can implement numerical computations based on equation (6). Partial results are tabulated as below (Table 4). The iteration results are generally consistent with the observational values of fractal dimension and the corresponding predicted values of the empirical models (Figure 3).

**Table 4 Comparison between observational values of Beijing's fractal dimension, predicted values of quadratic logistic models, and calculated values of 1-dimensional quadratic logistic map**

| Year | Capacity dimension | | | Information dimension | | | Correlation dimension | | |
|---|---|---|---|---|---|---|---|---|---|
| | $D_0$ | $\hat{D}_0$ | $\check{D}_0$ | $D_1$ | $\hat{D}_1$ | $\check{D}_1$ | $D_2$ | $\hat{D}_2$ | $\check{D}_2$ |
| **1984** | 1.4994 | 1.5003 | 1.4994 | 1.3961 | 1.4082 | 1.3961 | 1.3553 | 1.3737 | 1.3553 |
| **1988** | 1.5091 | 1.5204 | 1.5246 | 1.4234 | 1.4295 | 1.4223 | 1.3935 | 1.3952 | 1.3817 |
| **1989** | 1.5374 | 1.5313 | 1.5368 | 1.4467 | 1.4412 | 1.4351 | 1.4105 | 1.4069 | 1.3945 |

| | | | | | | | | |
|---|---|---|---|---|---|---|---|---|
| **1991** | 1.5645 | 1.5595 | 1.5674 | 1.4726 | 1.4712 | 1.4669 | 1.4392 | 1.4371 | 1.4266 |
| **1992** | 1.5849 | 1.5762 | 1.5853 | 1.5078 | 1.4890 | 1.4857 | 1.4797 | 1.4551 | 1.4455 |
| **1994** | 1.5944 | 1.6140 | 1.6251 | 1.5213 | 1.5292 | 1.5275 | 1.4920 | 1.4956 | 1.4878 |
| **1995** | 1.6522 | 1.6344 | 1.6464 | 1.5707 | 1.5509 | 1.5501 | 1.5373 | 1.5174 | 1.5106 |
| **1996** | 1.6789 | 1.6554 | 1.6682 | 1.5934 | 1.5733 | 1.5732 | 1.5560 | 1.5399 | 1.5340 |
| **1998** | 1.6702 | 1.6981 | 1.7122 | 1.5871 | 1.6186 | 1.6201 | 1.5531 | 1.5856 | 1.5816 |
| **1999** | 1.7174 | 1.7192 | 1.7337 | 1.6379 | 1.6409 | 1.6431 | 1.6056 | 1.6081 | 1.6050 |
| **2001** | 1.7423 | 1.7594 | 1.7743 | 1.6633 | 1.6834 | 1.6868 | 1.6308 | 1.6507 | 1.6495 |
| **2006** | 1.8559 | 1.8403 | 1.8530 | 1.7811 | 1.7675 | 1.7724 | 1.7462 | 1.7350 | 1.7370 |
| **2009** | 1.8667 | 1.8721 | 1.8821 | 1.7961 | 1.7998 | 1.8044 | 1.7647 | 1.7672 | 1.7699 |

**Note**: $D$ represents observational data, $\hat{D}$ represents predicted value of quadratic logistic models, and $\check{D}$ represents calculated value of 1-dimensional quadratic logistic map.

The results lend further support to the parametric models presented in Section 2 and empirical estimated values of the model parameters in Section 3. If the mathematical structure of a model is wrong, or if the parameter values of the model depart reality significantly, the calculated values will not keep consistent with the observed values. In this numerical simulation, the model is based on the quadratic logistic function, the initial values come from the observational data of fractal dimension, and the maximum fractal dimension values result from the empirical analysis. Besides, more numerical experiments can be made as follows. First, discretizing the quadratic Boltzmann's equation yields another 1-dimensional map of fractal dimension growth. Using this map, we can implement another numerical simulation. The new computational results are very similar to the results shown in Figure 3 and Table 4, and thus are no longer displayed here. Second, using the 2-dimensional maps, equations (19) and (20), we can also carry out a two-variable numerical experiment to verify the quadratic logistic models of fractal dimension growth curves. The process is similar to the process of generating urbanization level curves using the 2-dimensional maps based on the urban-rural interaction model (Chen, 2009).

### 3.4 Predictions and measurements of urban growth

By means of the models of fractal dimension growth curves, we can predict the peak and velocity of urban growth. After all, one of the main functions of mathematical model is prediction (Fotheringham and O'Kelly, 1989; Kac, 1969). The growth rate of fractal dimension values of urban form reflects the expansion speed of urban land use or the filling speed of urban space, and thus reflect the speed of urban human activities. For the quadratic logistic model, the growth rate of fractal dimension values can be expressed by equation (7), which is the derivative of equation (5). Discretizing equation (7) yields a difference equation of fractal dimension growth rate as below:

$$\frac{\Delta D_t}{\Delta t} = D_{t+1} - D_t = 2k^2 t D_t (1 - \frac{D_t}{D_{max}}), \quad (24)$$

in which the time difference is $\Delta t=1$. Equation (24) is equivalent to the 1-dimensional quadratic logistic map, equation (6). Using equation (24), we can generate the three time series of growth rates of fractal dimension. The parameter $k$ values are shown in Table 2, that is, 0.0626 for capacity dimension, 0.0639 for information dimension, and 0.0642 for correlation dimension. However, in practice, we have simpler way to create the time series of growth rates of fractal dimension for Beijing's urban form. The first step is to assign parameters to the quadratic logistic models. This step has been completed in Subsection 3.2. The second step is to calculate the predicted values of fractal dimensions using the models shown in Subsection 3.2. Partial values are displayed in Table 4. The third step is to compute the difference values of the fractal dimension time series, and the difference series represents the growth rates of fractal dimension values. The fourth step is to draw the plots of fractal dimension growth rate. Through these curves of growth rates, we can intuitively examine the regularity and characteristics of fractal dimension changes of Beijing's urban form (Figure 4).

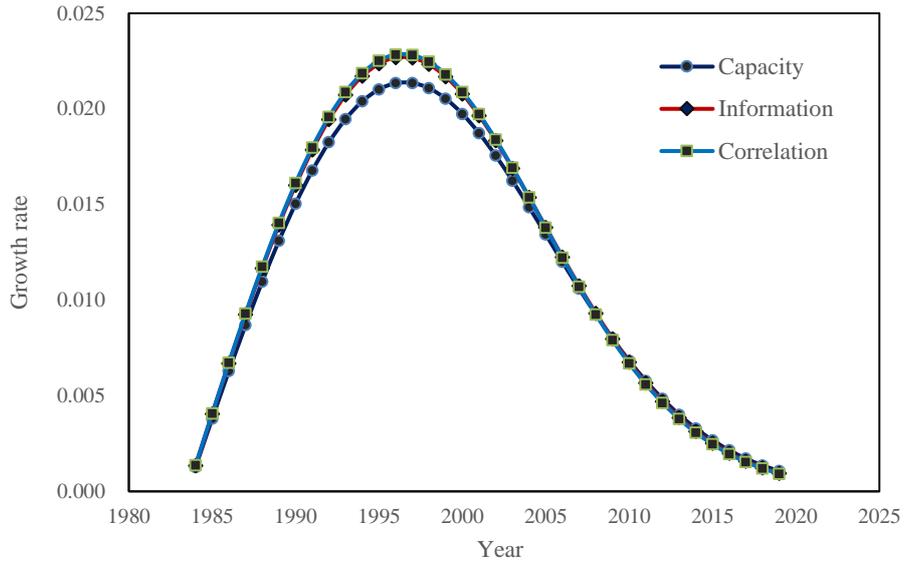

a. Based on quadratic logistic model

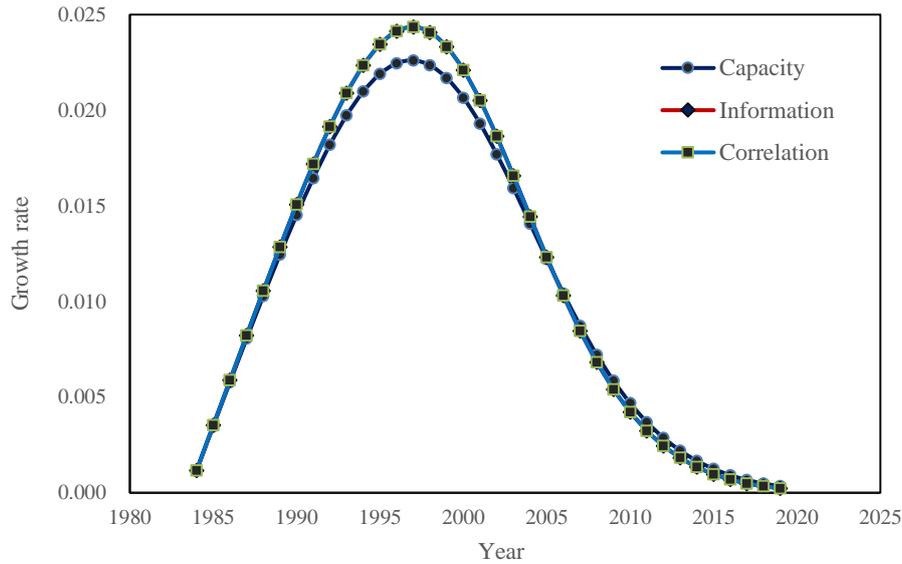

b. Based on quadratic Boltzmann model

**Figure 4 Two sets of unimodal curves reflecting the growth rate of fractal dimension values of Beijing's form (1984-2020)**

**Note**: According to the study area defined in this paper, the peak of urban land use appeared around 1997. The information dimension and correlation dimension curves cannot be differentiated in the two plots.

The empirical analyses of this section are mainly based on the quadratic logistic model. The reasons are as follows. First, the quadratic logistic model is simpler and easy to operate. Second, as indicated above, the fractal dimension growth curves are obtained by means of fixed study area. Actually, we can use the quadratic Boltzmann equation to predict urban growth rates. The results are similar to those from the quadratic logistic model, and the conclusion is the same as that from the quadratic logistic model. According to the rate curves of fractal dimension growth based on the quadratic logistic model, the rate peaks appeared between 1996 and 1998 years. The rate peak of capacity dimension came between 1997 and 1998, and the rate peaks of information and correlation dimensions came between 1996 and 1997 (Figure 4(a)). According to the rate curves of fractal dimension growth based on the quadratic Boltzmann model, all the peaks of capacity, information, and correlation dimensions appeared between 1997 and 1998 years (Figure 4(b)). Thus, a conclusion can be reached that the climax of land use of Beijing city took place about 1997 years. In other words, in between 1996 and 1998, the pace of urban space filling and expansion in Beijing reached its extreme.

As indicate above, due to scale-dependence of conventional urban measurements, fractal dimension is employed to act as a characteristic parameter of space filling to replace urban area in urban studies. So far, a recognized method has not been found to determine the urban boundary objectively. In particular, the fractal dimension values depend on the definition of study area owing to multifractal property of urban morphology. As a result, for the same city based on different study area, the rate peak of urban development may appear in different times. Therefore, before making use of the models of fractal dimension growth curves to predict urban growth, we must make clear the definition of study region: based on *city proper* (CP), based on *urbanized area* (UA), or based on *metropolitan area* (MA), and so on.

## 4. Discussion

Using experiment method, we establish mathematical models of fractal dimension growth curves of urban form. The basic aim of modeling is to predict the urban growth and explain urban dynamics in China. The delicate expression of the model is the quadratic Boltzmann equation, while the simple expression is the quadratic logistic function. We agree the following viewpoint: one of the main tasks of scientific research is to make models (Neumann, 1961). A model is different from the truth. A model's expression of a system is not unique. We can find a better model for a system, but maybe we will never find the best model for the system. If and only if a model can be derived from one or a set of postulates by mathematic theory, the model will evolve into a theoretical model from the empirical model. To derive the quadratic logistic model, we propose a group of spatial measurement and construct a pair of 2-dimensional dynamic equations. The models are based on Beijing's fractal dimension data. However, the results of modeling can be applied to many other Chinese cities. Setting $D_{min}$=0, Zhao (2017) fitted equation (21) to fractal dimension time series of the 13 main cities in Beijing, Tianjin, and Hebei region (Jing-Jin-Ji). The results show that the fractal dimension growth curves of these cities can be described with the various logistic functions: one logistic model (Baoding city, $b≈1$), two fractional logistic model (Shijiazhuang and Tianjin, $b=1.5$), and ten quadratic logistic model (Beijing, Cangzhou, Chengde, Handan, Hengshui, Langfang, Qinhuangdao, Tangshan, Xingtai, Zhangjiakou, $b=2$). All these cities belong to the northern China. A noticeable discovery is that although the definition of city or study area (e.g., CP, UA, MA) influences fractal dimension values and model parameter values of fractal dimension growth curves, it has no significant impact on the model's expression (Table 5). A model's structure reflect the spatial order and pattern at the macro level of a system, while the parameter values are often associated with the interaction between elements at the micro level. Another discovery is that the fractal dimension growth curves of the cities in the south of China such as Hangzhou, Shenzhen, and Yiwu can be described with the ordinary logistic function rather than the quadratic logistic function. The reason may be that, compared with northern Chinese cities, the development of southern Chinese cities is more significantly acted by the market economy of bottom-up evolution. It is high-level or large-scale macro factors that determine the mathematical structure of a city's models, whereas the model parameters are mainly affected by the micro interaction of urban internal elements.

**Table 5 A set of examples of mathematical models of fractal dimension growth curves of Chinese urban form: logistic function, quadratic logistic function, and fractional logistic function**

| City | Study area | Period | Data point | Mathematical model | $R^2$ |
|---|---|---|---|---|---|
| Beijing | Urban agglomeration | 1984-2009 | 13 | $\hat{D}_0(t) = \dfrac{1.9171}{1+0.2778e^{-(0.0626t)^2}}$ | 0.9811 |
| Beijing | Metropolitan area | 1984-2009 | 8 | $\hat{D}_0(t) = \dfrac{1.8627}{1+0.3070e^{-(0.0572t)^2}}$ | 0.9955 |
| Jing-Jin-Ji | Urban region | 1995-2013 | 5 | $\hat{D}_0(t) = \dfrac{1.5641}{1+0.0788e^{-(0.0292t)^2}}$ | 0.9783 |
| Shijiazhuang | Administrative area | 1995-2013 | 5 | $\hat{D}_0(t) = \dfrac{2}{1+0.4128e^{-(0.0370t)^{1.5}}}$ | 0.9428 |
| Tangshan | Administrative area | 1995-2013 | 5 | $\hat{D}_0(t) = \dfrac{1.5289}{1+0.1246e^{-(0.0402t)^2}}$ | 0.9428 |
| Tianjin | Administrative area | 1995-2013 | 5 | $\hat{D}_0(t) = \dfrac{1.7024}{1+0.1257e^{-(0.0549t)^{1.5}}}$ | 0.9355 |
| Shenzhen | Built-up area | 1986-2017 | 6 | $\hat{D}_0(t) = \dfrac{1.8636}{1+0.1950e^{-0.1101t}}$ | 0.9844 |

**Note**: Jing-Jin-Ji represents the "Beijing Tianjin Hebei region" in the North China Plain. Tianjin, Shijiazhuang, Tangshan, and Jing-Jin-Ji's data come from Zhao (2017), Shenzhen's data come from Ms. Xiaoming Man.

The models of fractal dimension growth curves for the urban form in China are different from those for the cities in western developed countries and regions. The growth curves of fractal dimension of European and American urban form can be modeled with the ordinary logistic function (Chen, 2012; Chen, 2014a; Chen, 2018). Benguigui *et al* (2000) have published several sets of fractal dimension time series (1935-1991) of the morphology of the Tel-Aviv metropolis (7-8 years), and Shen (2002) has presented a fractal dimension time series (1792-1992) of Baltimore's urban form (12 years). All these fractal dimension time series can be fitted by the common logistic function (Chen, 2012; Chen, 2018). Sun and Southworth (2013) have displayed the fractal dimension time series (1986-2010) of the cities and towns in the developed areas of Amazon tri‑national frontier regions (6 years). All the fractal dimension sets but one can be described with the logistic function, while the exception is suitable for the quadratic logistic function (Chen, 2014a; Chen, 2018). What is more, Murcio *et al* (2015) have calculated the multifractal parameter time series (1786-2010) of London (9 years), and the growth curves of capacity dimension, information dimension, and correlation dimension accord with the ordinary logistic curve rather than quadratic logistic curve. The sigmoid curves of fractal dimension growth of a country seems to be consistent with urbanization curves. The growth curves of urbanization level of developed countries is described with the conventional logistic function (Cadwallader, 1996; Chen, 2009; Karmeshu, 1988; Pacione, 2005; United Nations, 1980). However, in China and India, urbanization curves are consistent with the quadratic logistic curve (Chen, 2014b; Chen, 2016). The sigmoid curves of urbanization and fractal dimension increase indicate replacement dynamics (Chen, 2012; Chen, 2014a; Rao, 1989). Different logistic models of fractal dimension growth suggest different spatial replacement dynamics (Figure 5). An urban-rural interaction model is presented to explain the logistic growth of fractal dimension of the cities in developed countries (Chen, 2012; Chen, 2018). In this paper, another urban-rural coupling model is proposed for the cities in China, and this model can be used to describe the spatial dynamics of Chinese cities and derive the quadratic logistic model of fractal dimension growth curves. The similarities and differences of fractal dimension growth and the related dynamics can be tabulated as follows (Table 6).

**Table 6 A comparison of models of fractal dimension growth curves and the related dynamics between Chinese cities and the cities in western countries**

| Item | Cities in western countries (e.g. London) | Cities in China (e.g. Beijing) |
|---|---|---|
| Curve | S-shaped curve of odd symmetry | S-shaped curve of asymmetry |
| Boltzmann equation | $D(t) = D_{\min} + \dfrac{D_{\max} - D_{\min}}{1 + [\dfrac{D_{\max} - D_{(0)}}{D_{(0)} - D_{\min}}]e^{-kt}}$ | $D(t) = D_{\min} + \dfrac{D_{\max} - D_{\min}}{1 + [\dfrac{D_{\max} - D_{(0)}}{D_{(0)} - D_{\min}}]e^{-(kt)^2}}$ |
| Standard growth model | $D^*(t) = \dfrac{1}{1+(1/D^*_{(0)}-1)e^{-kt}}$ | $D^*(t) = \dfrac{1}{1+(1/D^*_{(0)}-1)e^{-(kt)^2}}$ |
| General growth model | $D(t) = \dfrac{D_{max}}{1+(D_{max}/D_{(0)}-1)e^{-kt}}$ | $D(t) = \dfrac{D_{max}}{1+(D_{max}/D_{(0)}-1)e^{-(kt)^2}}$ |
| 1-D dynamic model | $\dfrac{dD(t)}{dt} = kD(t)[1-\dfrac{D(t)}{D_{max}}]$ | $\dfrac{dD(t)}{dt} = 2k^2 tD(t)[1-\dfrac{D(t)}{D_{max}}]$ |
| 2-D dynamic model | $\begin{cases}\dfrac{dU(t)}{dt} = \alpha U(t) + \beta \dfrac{U(t)V(t)}{U(t)+V(t)} \\ \dfrac{dV(t)}{dt} = \lambda V(t) - \beta \dfrac{U(t)V(t)}{U(t)+V(t)}\end{cases}$ | $\begin{cases}\dfrac{dU(t)}{dt} = t[\alpha U(t) + \beta \dfrac{U(t)V(t)}{U(t)+V(t)}] \\ \dfrac{dV(t)}{dt} = t[\lambda V(t) - \beta \dfrac{U(t)V(t)}{U(t)+V(t)}]\end{cases}$ |
| Social mechanism | Process of bottom up urban evolution under top down rules (self-organized cities) | Interweaved processes of top down development and bottom up evolution of cities (limited self-organized cities) |

**Note**: The standard growth model is based on standardized fractal dimension values, while the general growth model is based on normal fractal dimension values. The 1-D dynamic models give the growth rates of fractal dimension of urban form and can be termed growth velocity models of fractal dimension.

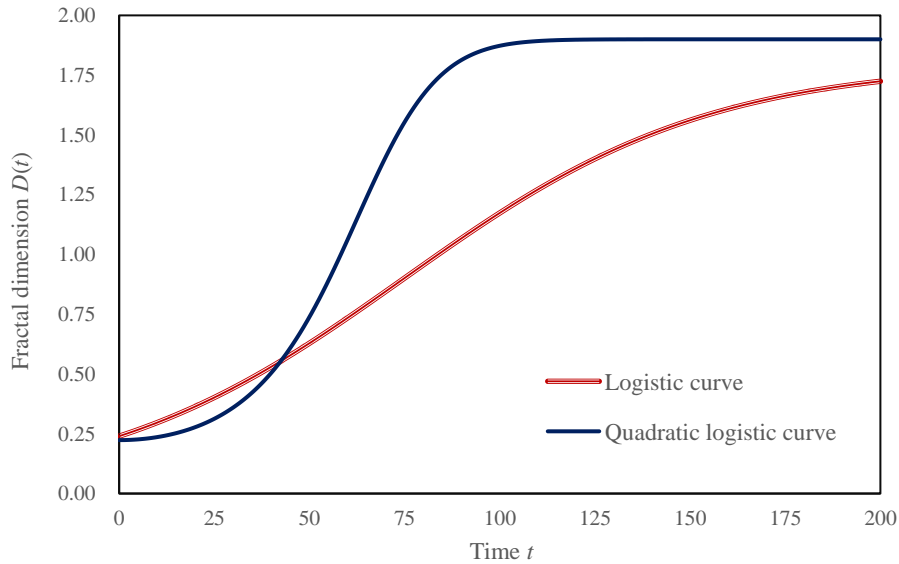

**Figure 5 Two types of fractal dimension growth curves of urban form: logistic curve and quadratic logistic curve**

**Note**: The logistic curve is based on London's model (Chen, 2018), while the quadratic logistic curve is based on Beijing's model. The former is a standard S-shaped curve of odd symmetry, while the latter is an S-shaped curve of asymmetry. The different curves suggests different types of spatial dynamics based on different social conditions.

The distinction between the models of the cities in western countries and those in China lies in latent scaling parameter value. The mathematical forms of the models presented in this paper look like the expressions put forward in a previous paper (see Chen, 2012). However, the essence is different. The previous models are common sigmoid functions (the scaling parameter is 1) for describing the fractal dimension growth curves of urban form in developed countries. In contrast, this work is to develop quadratic sigmoid function (the scaling parameter is 2) for characterizing the fractal dimension change of urban growth in China. Sometimes, the latent scaling exponent comes between 1 and 2. The models of spatial dynamics are significantly distinct from one another. For the previous model, the rates of growth of filled and unfilled space are independent of time, i.e., there is no dominant time factor on the right side of the equal signs of dynamic equations. However, in the new model, the growth rates of filled and unfilled space depend directly on time.

Fractal dimension is a substitute of urban area, and fractal dimension values reflect the expansion of urban agglomerations and space filling of urbanized area. According to the law of allometric growth of urban size and shape, urban area extension is correlated with population size growth. Population is one the central variables in the study of spatial dynamics of city development (Dendrinos, 1992), and it represent the first dynamics of urban evolution (Arbesman, 2012). Urban population activities are imprinted onto the form of urban land use, and what we measure and calculate is the fractal dimension of urban land use. Generally speaking, there is an allometric scaling relation between urban area and the corresponding population size (Batty and Longley, 1994; Chen, 2014b; Lee, 1989; Longley *et al*, 1991). Where Beijing is concerned, the urban area shown in Table 1 can be approximately modeled using the quadratic logistic function as below

$$\hat{A}(t) = \frac{2847.8631}{1+4.8828e^{-(0.0817t)^2}}.$$

The goodness of fit is about $R^2$=0.9886. Strictly, the more advisable scaling exponent for urban area model is $b$=3/2, thus we have fractional logistic model rather than a quadratic logistic model. The parameter values are as follows $A_{max}$=3213.7225, $a$=6.6329, $k$=0.0914. The goodness of fit is about $R^2$=0.9923. For simplicity and comparability, the latent scaling factor is approximated as 2. Accordingly, the population size growth of Beijing city can also be fitted using the quadratic logistic model (Appendix 2). However, if we examine the data of Beijing's urban population data, we will find that the better model is the hyperbolic function rather than the generalized logistic function. The hyperbolic model is in expression similar to equation (23). The hyperbolic growth model suggests that Beijing's population growth is very fast and lacks effective environmental constraints. Fast population urbanization leads to a rapid increase in fractal dimension values.

The main shortcomings of this study are as below. First, only one Chinese city is examined in detail. If we investigate more cities in developing countries, we may find more problems for studies in next step. Second, the relationships between fractal dimension and the measurements of space-filled extents in urban and rural regions are not brought to light. The more effective measurements for urban and rural space may be spatial entropy, which will be studied in a companion paper. Third, the numerical calculations based on the 2-dimensional map are not fulfilled. Owing to absence of urban and rural space-filling measurements, the numerical simulation of fractal dimension growth cannot be implemented by the discrete form of nonlinear coupling equations.

## 5. Conclusions

The first type of prediction models of fractal dimension growth include Boltzmann equation and logistic function. This work is devoted to making the second type of parametric models of growth curves of fractal dimension of urban form by means of experimental method. Using this type of models, we can predict the fractal dimension growth of many Chinese cities. Based on the theoretical modeling, empirical analysis, and discussion of questions, the main conclusions of this study can be reached as follows. **First, the fractal dimension growth curve of urban form in China can be modeled by quadratic Boltzmann's equation.** If a fractal dimension time series is normalized, the quadratic Boltzmann's equation will change to a quadratic logistic function. Both the quadratic Boltzmann's equation and quadratic logistic function belong to the family of quadratic sigmoid functions. For simplicity, the quadratic logistic function can be used to approximately describe the sample path of the fractal dimension series values that is not normalized in practice. **Second, the quadratic logistic functions of the fractal dimension growth curves of urban form can be derived from the equations of 2-dimensional spatial dynamics.** If urban space is divided into two parts: filled space (used space) and unfilled space (remaining space), we can build a spatial interaction model of the two types of geographical space. The model can be formulated as two differential equations indicative of spatial dynamics of urban evolution. From the pair of differential equations, we can deduce the sigmoid functions of urban fractal dimension growth curves. **Third,** **the models of the urban growth and form in China are the same in structure as the related models of the cities in western countries, but the model parameter values are different in essence.** As far as the mathematical expressions are concerned, the models of fractal dimension growth curves of Chinese cities are very similar to those of western cities. However, where model parameter values are concerned, there is significant difference. The temporal scaling exponent in the models of the fractal dimension growth curves of the cities in western countries is 1, while the scaling exponent value of Chinese cities are greater than 1 and often close to 2. The difference of parameter values reflects the distinction of dynamic mechanism at the micro level. The time factor is much more significant in the evolution of Chinese cities than in the urban evolution of western developed countries. Differences in time scaling factor values reflect differences in social and economic management systems between China and the western countries.

## Acknowledgements

This research was sponsored by the National Natural Science Foundations of China (Grant No. 41590843 & 41671167). The supports are gratefully acknowledged. We are very grateful to five anonymous reviewers whose constructive suggestions were helpful in improving the paper's quality.

# Appendices

## 1. New space-filling measurements of city development

A possible space-filling measure is the ratio of actual urban area to the maximum urban area. However, this ratio is not in direct proportion to fractal dimension or the ratio of actual fractal dimension to the maximum fractal dimension. An advisable measurement is the logarithm of urban area, that is, $U_t=\ln A_t$, which is similar to the definition of spatial entropy of urban land use. Thus the normalized space-filling index is

$$U_t^* = \frac{\ln A_t}{\ln A_{\max}}. \tag{A1}$$

If so, we will have an approximate relation as below

$$\frac{\ln A_t}{\ln A_{\max}} \approx \frac{D_t}{D_{\max}}. \tag{A2}$$

That is, $U_t^* \to D_t^*$. Empirically, the maximum urban area and maximum fractal dimension can be estimated using proper logistic models. Where Beijing is concerned, the $A_{\max}$ value can be estimated using the quadratic logistic model and the data in Table 2, and the $D_{\max}$ value can be found in Table 3. Taking the capacity dimension as an example, we can compute the normalized space-filling index, which is close to the normalized fractal dimension (Table A). The rural space-filling index can be worked out using equation (11), that is $V_t = U_t(1/D^*-1) = \ln A_t(1/D^*-1)$.

Table A The normalized capacity dimension and the corresponding urban space measurement based on urban area

| Year | Urban area $A$ | Dimension $D_0$ | $U^*=\ln A/\ln A_{\max}$ | $D^*=D/D_{\max}$ | $V^*=1-D^*$ | $U^*/D^*$ |
|---|---|---|---|---|---|---|
| 1984 | 393.3022 | 1.4994 | 0.7399 | 0.7821 | 0.2179 | 0.9460 |
| 1985 | 538.0889 | 1.5091 | 0.7787 | 0.7871 | 0.2129 | 0.9892 |
| 1986 | 530.2158 | 1.5374 | 0.7769 | 0.8019 | 0.1981 | 0.9688 |
| 1987 | 644.8084 | 1.5645 | 0.8011 | 0.8161 | 0.1839 | 0.9816 |
| 1988 | 729.0497 | 1.5849 | 0.8163 | 0.8267 | 0.1733 | 0.9874 |
| 1989 | 867.1342 | 1.5944 | 0.8378 | 0.8316 | 0.1684 | 1.0074 |
| 1990 | 1032.6239 | 1.6522 | 0.8594 | 0.8618 | 0.1382 | 0.9972 |
| 1991 | 1073.5342 | 1.6789 | 0.8642 | 0.8757 | 0.1243 | 0.9869 |
| 1992 | 1087.2600 | 1.6702 | 0.8658 | 0.8712 | 0.1288 | 0.9938 |

| | | | | | |
|---|---|---|---|---|---|
| **1993** | 1417.9022 | 1.7174 | 0.8987 | 0.8958 | 0.1042 | 1.0032 |
| **1994** | 1576.4091 | 1.7423 | 0.9118 | 0.9088 | 0.0912 | 1.0033 |
| **1995** | 2349.7967 | 1.8559 | 0.9612 | 0.9681 | 0.0319 | 0.9929 |
| **1996** | 2660.2626 | 1.8667 | 0.9766 | 0.9737 | 0.0263 | 1.0030 |

**Note**: According to the quadratic models of urban growth and fractal dimension growth curves, the maximum urban area $A_{max}$=3213.7225, and the maximum capacity dimension $D_{max}$=1.9171.

## 2. Modeling population growth of Beijing city

The generalized logistic model can be fitted to the dataset of the population size growth of Beijing city. We have no urban population which strictly corresponds to the urban area. Four years of urban population based on census data are available, that is, 1982, 1990, 2000, and 2010 (Table B). The time period of these population data is generally in accord with the time period of urban area and fractal dimension. The number of data points is less, but we can examine the basic trend. Significantly, the urban population size of Beijing can be described with the quadratic logistic function rather than the ordinary logistic function. The model is

$$\hat{P}(t) = \frac{1852.7150}{1+2.8745e^{-(0.0588t)^2}},$$

in which $t=n$-1982, where $n$ denotes year. The goodness of fit is about $R^2$=0.9994. The observation values match the trend line very well.

**Table B Beijing's municipality population, city and town population, and urban population based on six times of census**

| Year | 1953 | 1964 | 1982 | 1990 | 2000 | 2010 |
|---|---|---|---|---|---|---|
| **Municipality population** | 276.8149 | 756.8495 | 923.0687 | 1081.9407 | 1356.9194 | 1961.2368 |
| **City and town population** | 205.8000 | 425.8000 | 597.0000 | 794.5000 | 1052.2000 | 1685.9000 |
| **City population** | -- | -- | 467.4921 | 576.9607 | 949.6688 | 1555.2378 |

**Note**: The municipality population is based on the concept of administrative area, including all the urban and rural population; the city and town population include the residents in central city and peripheral towns; the city population does not contain the residents in varied towns, and is generally consistent with urbanized area.